\documentclass[12pt]{article}
\usepackage{epsfig}

\date{}

\newcommand{\beeq}{\begin{equation}}
\newcommand{\eneq}{\end{equation}}
\newcommand{\be}{\begin{eqnarray}}
\newcommand{\ee}{\end{eqnarray}}
\newcommand{\bpic}{\begin{picture}}
\newcommand{\epic}{\end{picture}}

\def\la{\raise.16ex\hbox{$\langle$} \, }
\def\ra{\, \raise.16ex\hbox{$\rangle$} }

\def\psibar{ \psi \kern-.65em\raise.6em\hbox{$-$} }
\def\mbar{ m \kern-.78em\raise.4em\hbox{$-$}\lower.4em\hbox{} }

\def\n@space{\nulldelimiterspace=0pt \mathsurround=0pt }
\def\huge#1{{\hbox{$\left#1\vbox to 20.5pt{}\right.\n@space$}}}

\def\myskip{\noalign{\kern 8pt}}
\def\myeqspace{\noalign{\kern 10pt}}

\def\boxit#1{$\vcenter{\hrule\hbox{\vrule\kern3pt
    \vbox{\kern3pt\hbox{#1}\kern3pt}\kern3pt\vrule}\hrule}$}
\def\bigbox#1{$\vcenter{\hrule\hbox{\vrule\kern5pt
     \vbox{\kern5pt\hbox{#1}\kern5pt}\kern5pt\vrule}\hrule}$}

\def\ignore#1{{}}

\begin{document}

\bibliographystyle{unsrt}
\footskip 1.0cm

\thispagestyle{empty}
                           
\begin{flushright}
LBNL - 44316 Abs.\\
\end{flushright}

\vspace{1in}

\begin{center}{\Large \bf {Shadowing of Gluons at RHIC and LHC
\footnote{Talk given at XXIX Int. Symp. on Multiparticle Dynamics 
(ISMD99), 9-13 August 1999, Providence, RI, USA.}}}\\

\vspace{1in}
{\large  Jamal Jalilian-Marian}\\

\vspace{.2in}
{\it  Nuclear Science Division, Lawrence Berkeley National Laboratory, 
          Berkeley, CA, USA}\\

\end{center}

\vspace*{25mm}

\begin{abstract}
\baselineskip=18pt

We show estimates for shadowing of gluons at small values of
$x$, appropriate to RHIC and LHC experiments. Using a new evolution 
equation which takes into account the effects of gluon recombination
to all orders in gluon density, we show that there is a 
significant depletion in the gluon density of large nuclei.

\end{abstract}

\vspace*{5mm}

\newpage

\normalsize
\baselineskip=22pt plus 1pt minus 1pt
\parindent=25pt

\section{Introduction}

Understanding the initial conditions of a relativistic heavy
ion collision from first principles is perhaps the single 
most challenging problem facing the heavy ion community. Proposed
signatures of a possible Quark-Gluon Plasma (QGP) formed in
heavy ion collisions will crucially depend on the initial 
conditions. At the very early stages of the collision, one would 
need to take the full quantum mechanical nature of the nuclei
into account which is a prohibitingly difficult task since
it would require full knowledge of the nuclear wave functions.
The McLerran-Venugopalan model \cite{mv} offers a new and promising tool
to investigate these early times by representing the initial
nuclei by classical fields. At later times, when the 
highly off-shell modes of the field are freed by hard scattering
and go on-shell, one can identify these modes with partons
and consider the initial distributions of these partons in the nuclei. 

When calculating a typical nuclear cross section, one needs to
know the distribution of a given parton kind in the nucleus
$f_{g/A}$. Using the QCD factorization theorems, one can then
write the nuclear cross section as a convolution of the nuclear
parton densities with the hard parton-parton cross section, i.e.
\be
\sigma_{AB} \sim f_{g/A}(x,Q) \otimes \sigma_{gg} \otimes f_{g/B}(x,Q)  
\ee

In order to obtain the nuclear parton distributions, one can take
the corresponding parton distributions in a nucleon and scale them
by the atomic weight $A$. This sounds plausible specially at high
values of $Q^2$ since one does not expect nuclear effects to be 
important at large values of $Q^2$. However, this expectation was
proved to be too naive and it was experimentally found the the 
distribution of partons in free nucleons are strikingly different 
from those in bound nucleons. This difference is more pronounced 
in large nuclei at small values of $x_{bj}$ where a significant 
depletion in the number of partons is observed so that a simple 
$A$ scaling does not hold. 

Alternatively, one could measure the nuclear parton distribution
in a particular experiment. Since these distributions are universal,
one could then use them to predict nuclear cross sections in other
experiments. The most recent experiments measuring nuclear parton
distributions have been performed by the NMC collaboration at CERN
SPS \cite{arnetal} and by the E665 collaboration at 
Fermilab \cite{adametal}. However, both of these
are fixed target experiments and are limited in the kinematic range
in $x$ and $Q^2$ they can cover. Also, the amount of data 
in the kinematic region where 
perturbative QCD would apply is limited. A lepton-nucleus collider 
would go a long way towards expanding our knowledge of the nuclear 
parton distributions and is urgently needed.    

Once the nuclear parton distributions are known at a given value
$x_0$ and $Q_0$, one can use the perturbative QCD evolution equations
to predict the distributions at different $x$ and $Q$. However, 
the standard evolution equations are expected to break down at very 
small values of $x$ due to parton recombination effects. A new 
evolution equation (JKLW) which takes these effects into account was
derived in \cite{mnmob} and is the non-linear all twist generalization of the
standard perturbative QCD evolution equations such as DLA DGLAP
and GLR/MQ (see also \cite{aglyk} for a similar equation). These non-linear 
effects were investigated numerically 
in \cite{jw} and were found to be important in the kinematic region
to be explored by the upcoming experiments at RHIC and LHC. One can
also use this new evolution equation to predict the $x$, $Q$, $b_t$
and $A$ dependence of the gluon shadowing ratio defined as
\be
S={xG_{A} \over A xG_{N}}
\label{eq:shad}
\ee
where $xG_{A}$ and $xG_{N}$ are the nuclear and nucleon gluon
distribution functions respectively.

\section{Shadowing of gluons}

To calculate the shadowing ratio $S$ for gluons, we start with the 
following evolution equation which was derived from the 
effective action for QCD at small $x$ in \cite{mnmob}.
\be
{\partial^2 \over \partial y \partial \xi}\;xG(x,Q,b_\perp)=
\frac{N_c(N_c-1)}{2}\ Q^2\bigg[1 - 
{1 \over \kappa} \exp({1\over \kappa}) E_1({1\over \kappa})\bigg]
\label{eq:jklw}
\ee
where 
\be
\kappa={2 \alpha_s \over \pi (N_c -1)Q^2} xG (x,Q,b_\perp)
\ee
and ${\rm E_1}(x)$ is the exponential integral function defined 
as \cite{abr}
\be
{\rm E_1}(x)=\int_{0}^\infty\! dt\ {e^{-(1+t)x} \over 1+t},\ \ \ \ x>0
\ee 

This equation was shown to reduce to DLA DGLAP and GLR/MQ 
at the low gluon density limit. In \cite{jw} we showed in 
detail how to solve this equation numerically. Here, we briefly
review our main approximations and assumptions and refer the 
interested reader to \cite{jw} for more details. In order to solve
equation (\ref{eq:jklw}), we need to know the initial gluon
distribution at some reference point $x_0$ and $Q_0$ as well
as its derivative (this is due to making the semi-classical
approximation). We then use a fourth order Runge-Kutta code
to calculate the gluon distribution at any other point $x$ and $Q$.
In \cite{jw}, we took $x_0=0.05$ and $Q_0=0.7$, but the effective
$Q_0$ for most points calculated was about $1 GeV$. The reason
for our choices were two fold; first that experimentally it is known
that the shadowing ratio is about $1$ in the range $x=0.05-0.07$. Also,
in order to maximize the effects of perturbative shadowing, we needed
to start from as low value of $Q_0$ as possible while keeping it high
enough so that perturbative QCD is still valid. With these 
approximations, we showed in \cite{jw} that for large nuclei, the 
non-linearities of the evolution equation are very important. 
Our results for the gluon shadowing ratio as defined in (\ref{eq:shad})
are shown in Figure $1$. It is clear that gluon shadowing at
RHIC and LHC will be important.

\begin{figure}[ht]
\epsfxsize=40pc
\epsfbox{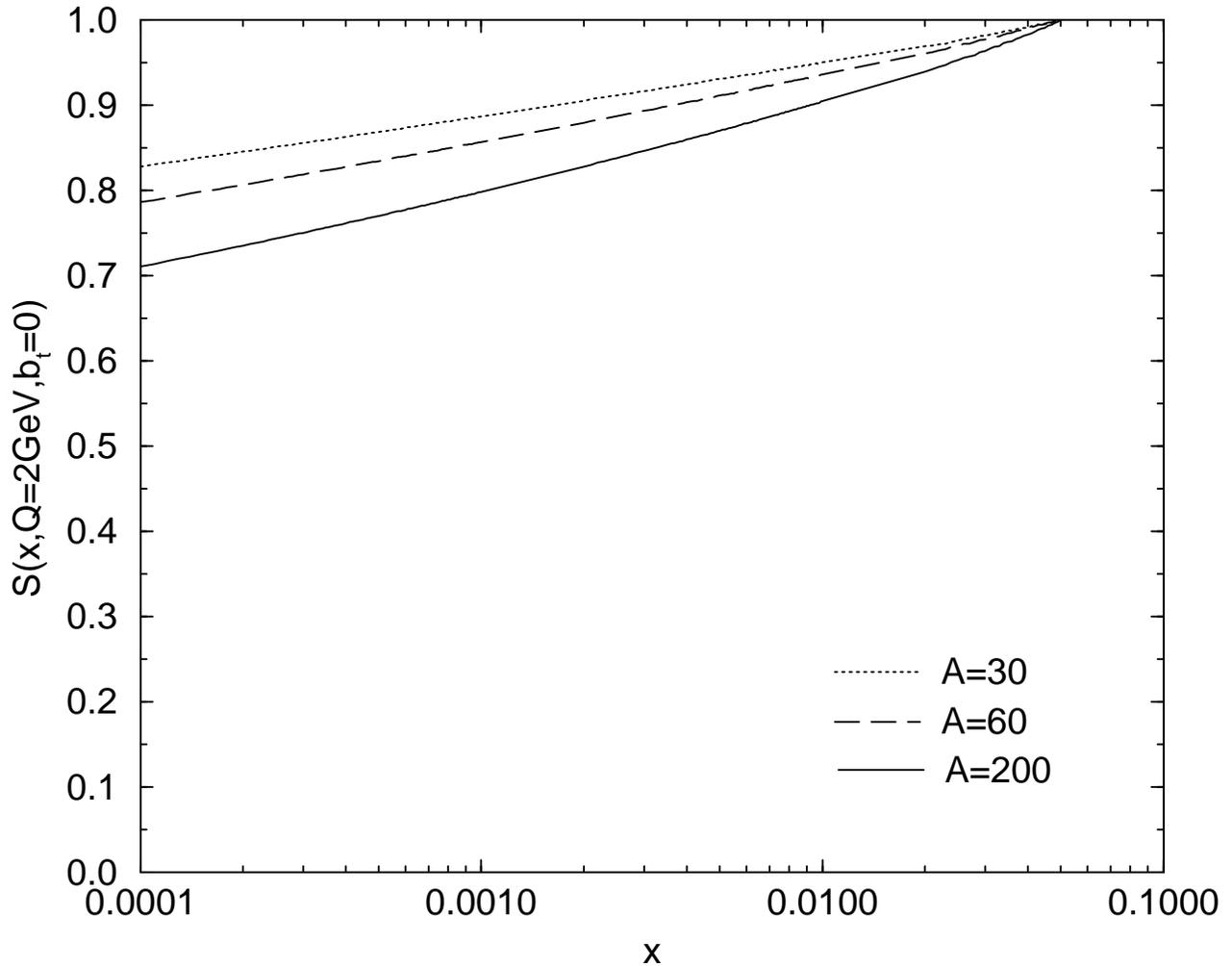}
\caption{Gluon shadowing as defined in (\ref{eq:shad})  
\label{fig:shadow}}
\end{figure}

Shadowing of gluons in nuclei will have significant effects
on the measured observables in the upcoming experiments at
RHIC and LHC. For example, initial minijet and total transverse
energy production will be greatly reduced. Also, heavy quark
production will be significantly effected since its cross section
is proportional to the squared of gluon density. Basically, any
production cross section which involves the distribution of gluons
in nuclei will be modified. Therefore, it is extremely important
to understand shadowing more thoroughly. For example, the observed
shadowing ratio (of $F_2$) does not seem to have a significant 
$Q$ dependence while parton recombination models tend to predict
a strong $Q$ dependence of the shadowing ratio. This could in
principle be due to assuming no shadowing at the initial point, 
i.e. the point where the perturbative evolution starts from. To
investigate this, one should include initial non-perturbative
shadowing at the reference point and then evolve the distributions
with both the leading twist DGLAP and all twist JKLW evolution
equations. The difference between the two would be a clear
indication of importance of higher twist effects in understanding
gluon shadowing.  
 
\leftline{\bf Acknowledgments} 

I would like to thank all of my collaborators and specially Xin-Nian 
Wang with whom much of the numerical work presented here has been done.
This work was supported by the Director, Office of Energy Research, 
Office of High Energy and Nuclear Physics Division of the Department of 
Energy, under contract No. DE-AC03-76SF00098 and DE-FG02-87ER40328.

\leftline{\bf References}

\renewenvironment{thebibliography}[1]
        {\begin{list}{[$\,$\arabic{enumi}$\,$]}  
        {\usecounter{enumi}\setlength{\parsep}{0pt}
         \setlength{\itemsep}{0pt}  \renewcommand{\baselinestretch}{1.2}
         \settowidth
        {\labelwidth}{#1 ~ ~}\sloppy}}{\end{list}}

\end{document}